\documentclass{iauc}
\usepackage{graphicx}
\pubyear{2005}
\volume{199}
\pagerange{1--}
\jname{Probing Galaxies through Quasar Absorption Lines}
\editors{P. R. Williams, C. Shu, and B. M\'{e}nard, eds.}

\title[Constraining MgII absorbers with the SDSS]{Constraining MgII absorber systems\\ with the SDSS}

\author[M\'enard, Zibetti, Nestor \& Turnshek] {Brice M\'enard$^1$,
Stefano Zibetti$^2$, Daniel Nestor$^3$ \and David Turnshek$^4$}

\affiliation{ $^1$Institute for Advanced Study, Einstein Drive,
Princeton NJ 08540, USA\break $^2$Max-Planck-Institut f\"ur
Extraterrestrische Physik, Garching, Germany\break $^3$Dept. of
Astronomy, University of Florida, Gainesville, FL 32611, USA\break
$^4$Dept. of Physics and Astronomy, University of Pittsburgh,
Pittsburgh, PA 15260, USA}

\def\e{et al.}
\def\MgII{MgII}
\def\arcsec{$^{\prime \prime}$}

\begin{document}

\maketitle

\begin{abstract} 
Using a large sample of MgII absorbers with $0.4<z<2.2$ detected
by Nestor \e\ (2005) in the Early Data Release of the SDSS, we present
new constraints on the physical properties of these systems based on
two statistical analyses: (i) By computing the ratio between the
composite spectra of quasars with and without absorbers, we measure
the reddening effects induced by these intervening systems; and (ii) by
stacking SDSS images centered on quasars with strong MgII absorption
lines and isolating the excess light around the PSF, we measure the
mean luminosity and colors of the absorbing galaxies.  This
statistical approach does not require any spectroscopic follow up and
allows us to constrain the photometric properties of absorber systems.
\end{abstract}

\firstsection 

\section{Introduction}

Since it was realized that strong metal absorption lines detected in
quasar spectra are due to intervening galaxies (e.g. Bahcall \&
Spitzer 1969, Boksenberg \& Sargent 1978, Bergeron \& Boiss\'e 1991),
significant progress has been made in our understanding of absorber
systems.  A number of studies have constrained some of their physical
properties such as temperature, abundances and kinematics (for a
review see for example Churchill's contribution). However, the precise
nature of absorber systems is still elusive and more observational
constraints are needed in order to improve our understanding on how
these systems trace various structures such as galactic disks, halos,
winds, etc. This would allow us to include and use absorbers in our global
picture of galaxy formation and evolution.

In this contribution we illustrate how imaging and spectroscopic data
from the Sloan Digital Sky Survey (SDSS) can be used to constrain the
nature of absorber systems. Having selected a population of quasars
with strong absorbers as well as a population of reference quasars (\S
2), we construct their composite spectra and isolate the reddening
effects due to the presence of absorbers (\S 3). In \S 4,
we show how the photometric properties of absorbing galaxies can be
constrained by measuring the statistical excess of light present
around quasars in SDSS images. We summarize the results in \S 5.

\section{The data}
\label{section_data}
\subsection{The MgII absorber sample}

\begin{figure}
\begin{center}
 \includegraphics[height=5cm,width=.45\hsize]{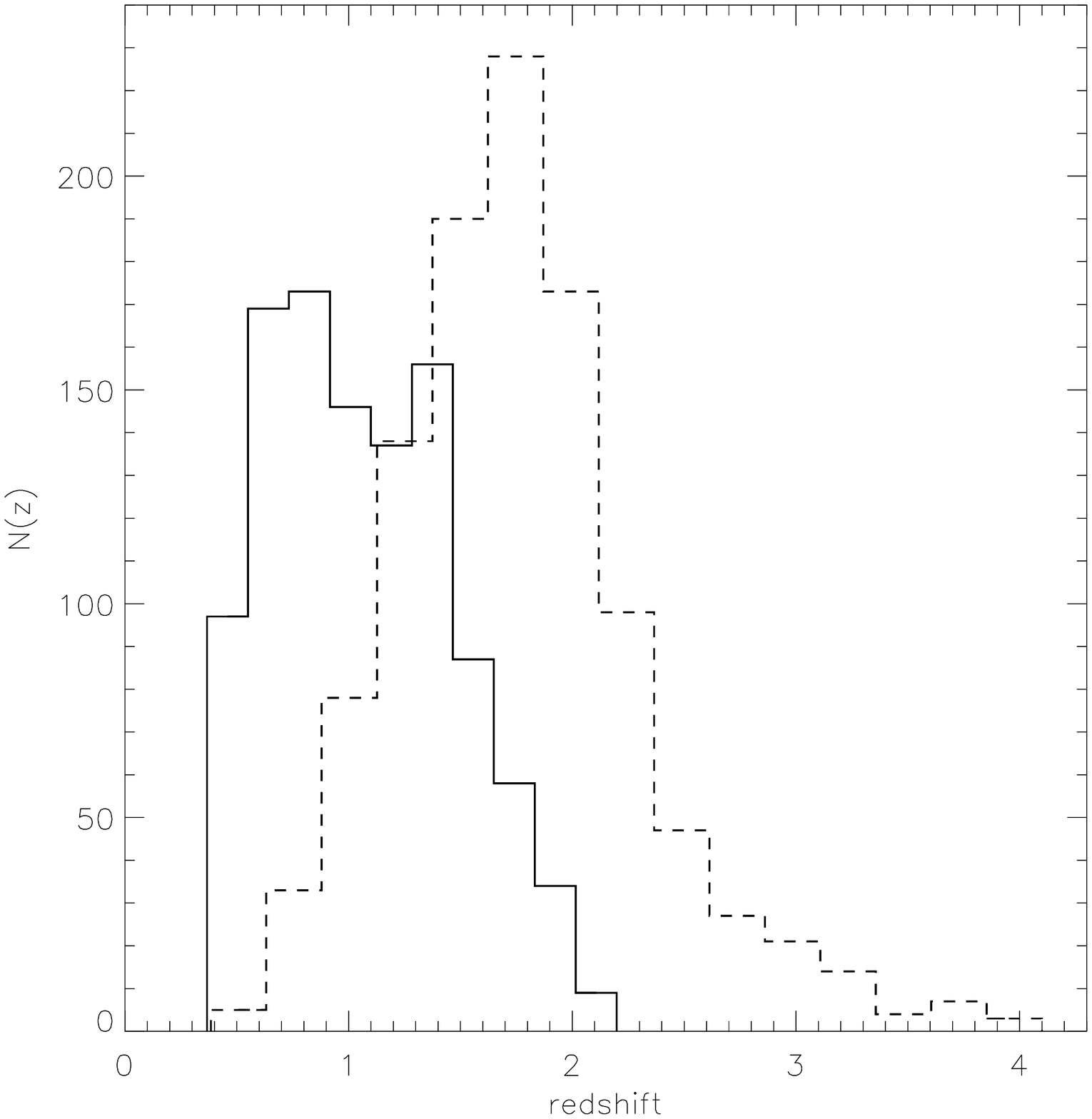}
 \includegraphics[height=5cm,width=.45\hsize]{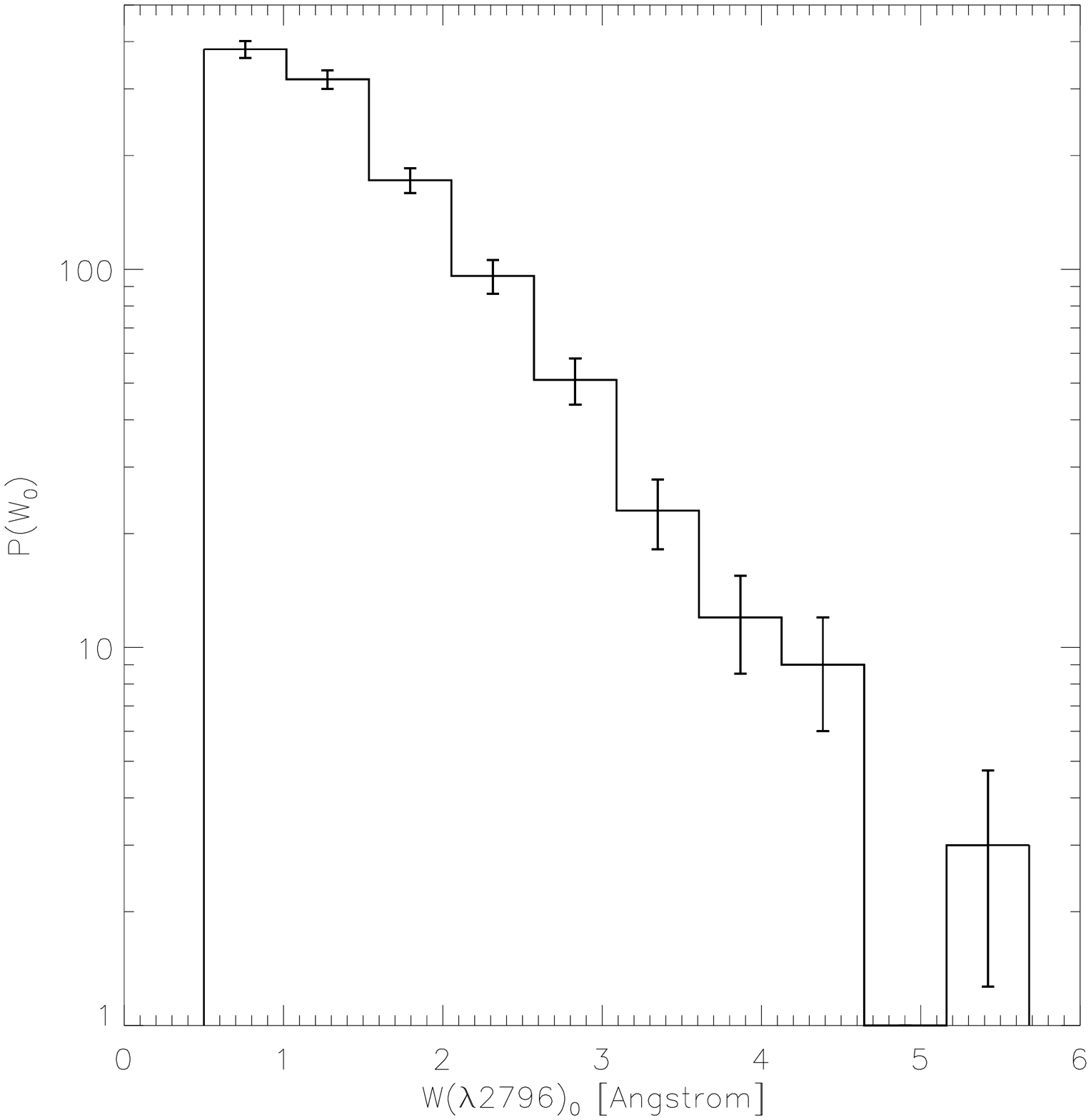}
  \caption{\emph{Left:} redshift distributions of the MgII absorbers
with $W(\lambda2796)_0>0.8$ \AA\ (solid line) and background quasars (dashed
line). \emph{Right:} rest equivalent width distribution of the
selected absorbers.}
\label{fig1} 
\end{center} 
\end{figure}

We use the sample of \MgII\ absorption line systems compiled by Nestor
\e\ (2005) based on SDSS EDR data (Stoughton et al. 2002). 
Here, we focus our analysis on strong systems
with a rest equivalent width $\mathrm{W_0(\lambda2796)}>0.8\;\mathrm{\AA}$.
In this range, the detection completeness is very high and multiple
systems along the same line-of-sight are rare. In addition, we 
focus our reddening measurements on the strongest systems, with
$\mathrm{W(\lambda2796)_0}>2.5\;\mathrm{\AA}$. In this section we briefly
summarise the main steps involved in the absorption line detection
procedure.  For details on the quasar and absorber catalogs, we refer
the reader to Schneider \e\ (2002) and Nestor \e\ (2005).

The SDSS EDR provides 3814 QSO spectra, approximately 3700 of which
are of QSOs with sufficiently high redshift to allow the detection
of intervening \MgII\ absorption lines.  The continuum-normalized SDSS
QSO spectra were searched for \MgII\ $\lambda\lambda2796,2803$
doublets using an optimal extraction method employing a Gaussian
line-profile to measure each rest equivalent width $W_0$. The
identification of MgII doublets required the detection of the 2796
line at more than 5$\sigma$ and one additional line detected at least
at 3$\sigma$. Only systems 0.1c blueward of
the quasar redshift and redward of Ly$\alpha$ emission were
accepted. Systems with separations of less than 500 km/s were
considered as a single system.  The final sample comprises 813 QSOs
with MgII absorbers.  We show the redshift distribution of the
absorbers with $\mathrm{W(\lambda2796)_0}>0.8$ \AA\ (solid line) and that of the
background quasars (dashed line) in the left panel of Figure
\ref{fig1}. The right panel shows the rest equivalent width
distribution $W_0$ of the 2796 \AA\ line. Among these systems, 55 have
$\mathrm{W(\lambda2796)_0}>2.5$~\AA.

\subsection{The reference quasar sample}
\label{section_selection}

In order to isolate the effects induced by the presence of absorber
systems we use a control sample made of quasars without absorbers.  To
avoid selection biases, the two quasar populations must have the same
redshift distribution and the same absorption-line detectability.  The
main steps of our object selection are as follows: for each quasar
with an absorber system, we randomly look for a quasar without any
detected absorber (down to the limiting value of the line finder,
i.e. $W(\lambda2796)_0=0.3\,\mathrm{\AA\ }$) such that their
redshifts are similar within 0.1. In addition, we require that this
quasar has a high enough S/N to allow the detection of absorption
lines of similar strength as the absorber system at wavelengths
corresponding to $z_{abs}$. Thus our procedure ensures that the lack
of absorption at $z_{abs}$ in the \emph{reference} QSO spectrum is
real.  The quasars selected in this way are called \emph{reference}
quasars in the following.  Contamination by undetected absorbers at
$z\neq z_{abs}$ is expected to statistically affect both samples in
the same way. We match each quasar with absorbers to two reference
quasars, which allows us to decrease the noise level.

\section{Reddening properties from composite spectra}

\begin{figure}
\begin{center}
 \includegraphics[height=7cm,width=.49\hsize]{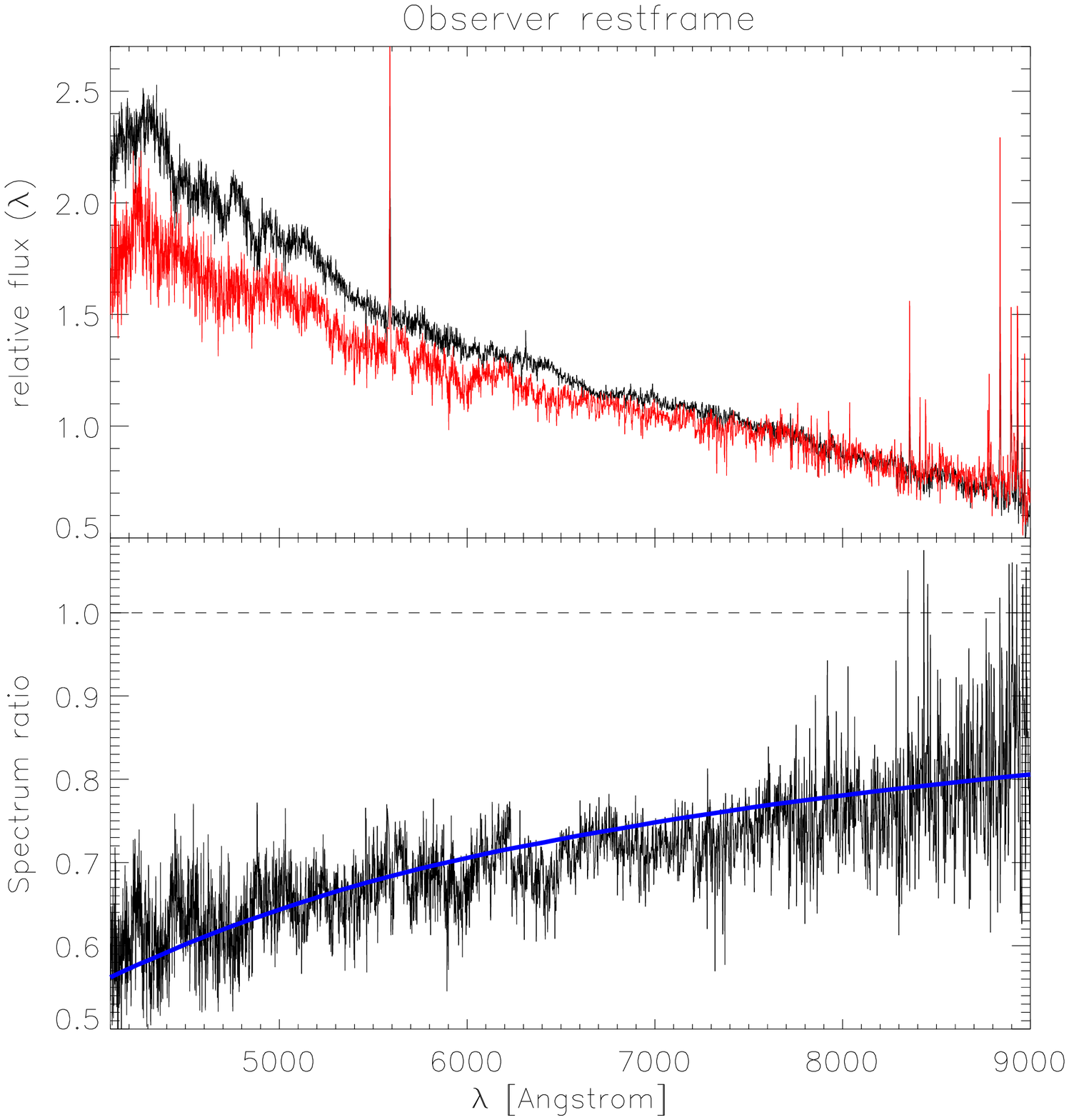}
 \includegraphics[height=7cm,width=.49\hsize]{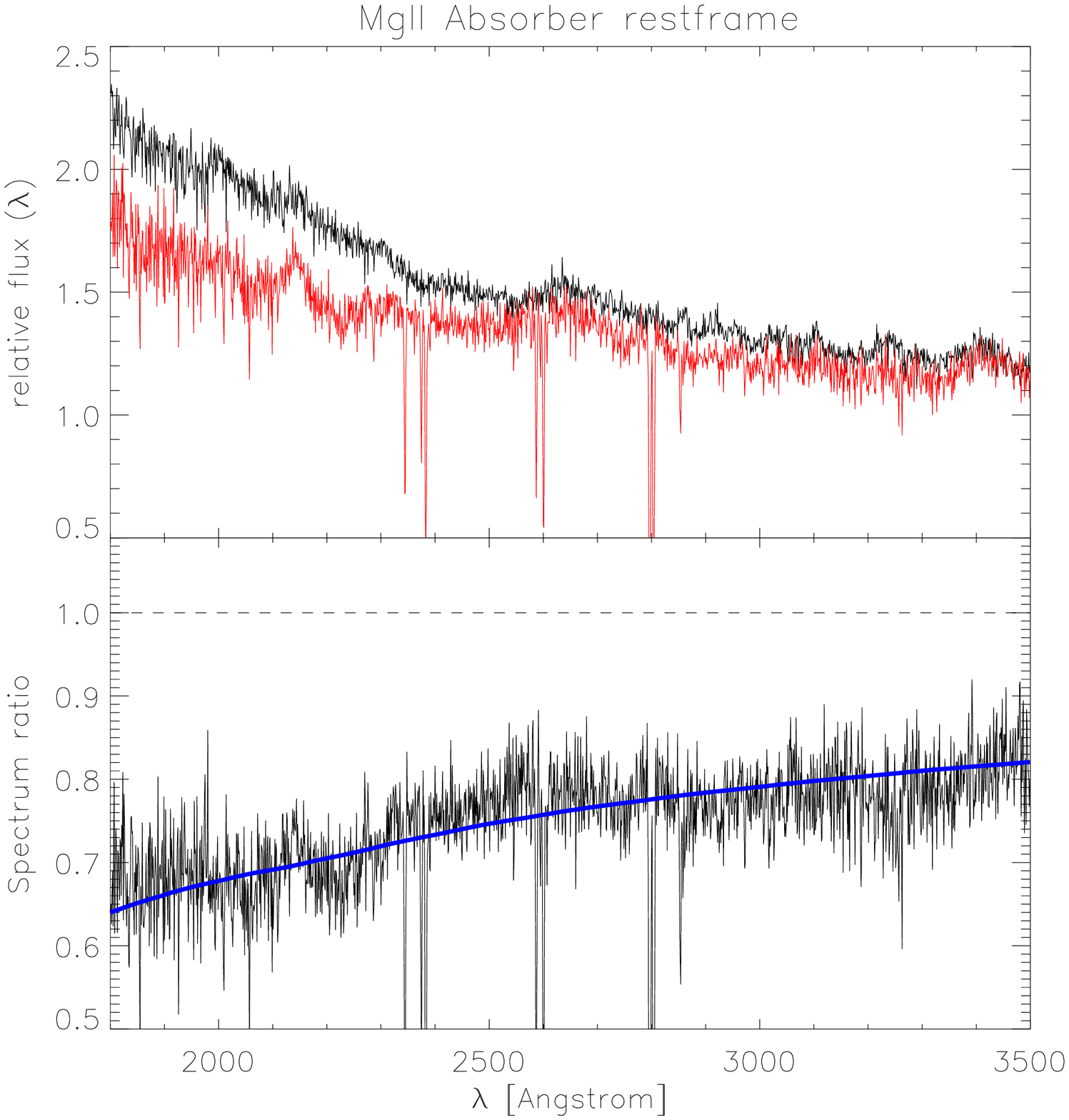}
  \caption{Composite spectra of quasars with and without strong MgII
absorbers with $W_0>2.5$\AA, in the observer (left) and absorber
(right) rest-frame. The lower panels show the ratio of the composite
spectra. The fitted curve used in the lower panels corresponds to the
SMC reddening curve.}  
\end{center} 
\label{fig2} 
\end{figure}

In this section we investigate the reddening and extinction properties
induced by our selected population of absorbers to their background
quasars. For each sample of quasars defined above we create composite
spectra in the observer- and absorber rest-frames. The composite
spectra are obtained by computing the geometric mean of all spectra
renormalized so that the global signal is not dominated by the
brightest objects and taking into account the masks provided by the
SDSS pipeline.  For computing the composite spectrum in the absorber
rest frame, sky lines are masked and the wavelength shifts applied to
the quasars with absorbers are also used for the reference
quasars. The results are shown in the upper panels of figure
\ref{fig2}.  The composite spectra are shown in the observer and
absorber rest-frame in the left and right panels. The corresponding
spectra have been normalised to the same flux at $\lambda=8500$ and
3500 \AA, respectively.  As can be seen in both panels, quasars with
absorbers appear to be redder than their reference quasars.  In order
to quantify this difference, we have computed the spectrum ratio and
displayed the result in the lower panels of the figure.  In the
absorber rest frame, this ratio provides a direct measurement of the
mean extinction curve of our sample of absorber systems.  In the
observer rest-frame, this quantity has been convolved with the
redshift distribution of the absorber systems.  As we can see, the
2175 \AA\ excess seen in the reddening curve of the Milky way or the
LMC is not present. We find that the SMC reddening curve provides a
good fit to the mean extinction curve of MgII absorbers\footnote{We
use the template provided by B. Draine at
http://www.astro.princeton.edu/\symbol{126}draine}. This is shown with
the smooth curve. In each case, we have normalized the ratio such that
it is unity at sufficiently large wavelengths. This allows us to
estimate the expected amount of extinction as a function of
wavelength.

Using the SMC reddening curve as a fitting template, the measured
reddening allows us to constrain the mean product $\langle
(k/k_{SMC})\,\mathrm{N_H}\rangle$ where $k_{SMC}$ is the SMC
dust-to-gas ratio and $\mathrm{N_H}$ is the hydrogen column density.
From the results obtained in the observer and absorber rest-frames, we
find $\langle (k/k_{SMC})\,\mathrm{N_H}\rangle=(2.59\pm
0.32)\times10^{21}$ and $(2.08\pm0.52)\times10^{21}$ atoms~cm$^{-2}$,
respectively.  Note that the noise contributions are different in each
case.  Our result indicates that MgII absorbers with
$W(\lambda2796)_0>2.5\,\mathrm{\AA}$ generally correspond to Damped
Ly$\alpha$ systems. We find the reddening values to strongly depend
on the MgII rest equivalent width. Detailed results will be presented
in Menard et al. (2005).

\section{Photometric properties from image stacking}

In this section we show that producing a high signal-to-noise (S/N)
\emph{average} image of the quasars with absorbers allows us to
measure the photometric properties of the absorbing galaxies.  For the
two samples of quasars defined in section 2, we stack the
corresponding SDSS images centered on the quasars. This procedure
consists of four basic steps: shifting to make images superposable,
rescaling the intensity to a uniform photometric calibration, masking
of all unwanted sources, and averaging.

We use the raw ``corrected frames'' publicly available in the Third
Data Release of the SDSS \cite[(Abazajian et al. 2005)]{dr3}.  The
image shift is done using a bi-cubic spline interpolant and the
photometric calibrations provided by the DR3 catalogs are used to
rescale the intensities.  We then generate masks blanking sources that
are likely to be foregrounds with respect to the absorber.  For each
QSO we create three independent masks in $g$, $r$ and $i$-band, that
are eventually combined in a final mask. In each band, we first
subtract from the corrected frame the Atlas Image of the QSO produced
by the \texttt{PHOTO} pipeline \cite[(Lupton et
al. 2001)]{Lupton01}. As the Atlas Image contains all and only the
pixels that are attributed to the QSO by the de-blender, we are left
with a clean image where all sources but the QSO are present. On this
image we run SExtractor \cite[(Bertin \& Arnouts 1996)]{Bertin06} to
find all candidate sources. We mask all objects with a flux exceeding
that expected for an extreme case of a luminous blue galaxy: a stellar
population with $\mathrm{M}_i<-22.0$ produced by a burst $10^7$ Gyr
long, observed immediately after the burst end at the redshift of the
absorber as predicted by \cite[Bruzual \& Charlot]{BC03}'s population
synthesis models (Bruzual \& Charlot 2003).  Our final results do not
critically depend on this choice of different spectral energy
distributions.  Finally, the masks in the three bands are combined and
conservative masks are applied to all objects classified as stars by
\texttt{PHOTO} with $r<20.0$. Non-saturated stars are masked out to
their 25 mag arcsec$^{-2}$ isophotal radius ($r$-band), while
saturated stars are masked out to three times this radius.

After subtracting the background value evaluated on the masked image,
the shifted and intensity-calibrated images are averaged, excluding
all masked pixels. Azimuthally averaged surface brightness profiles
are extracted in a series of circular apertures which are optimized to
integrate on larger areas at larger radii. A more accurate background
level for these profile is recomputed on the stacked image. The
complete covariance matrix of the extracted background-subtracted
surface brightness profile is evaluated using the jackknife technique.

\begin{figure}
\begin{center}
 \includegraphics[height=7cm,width=\hsize]{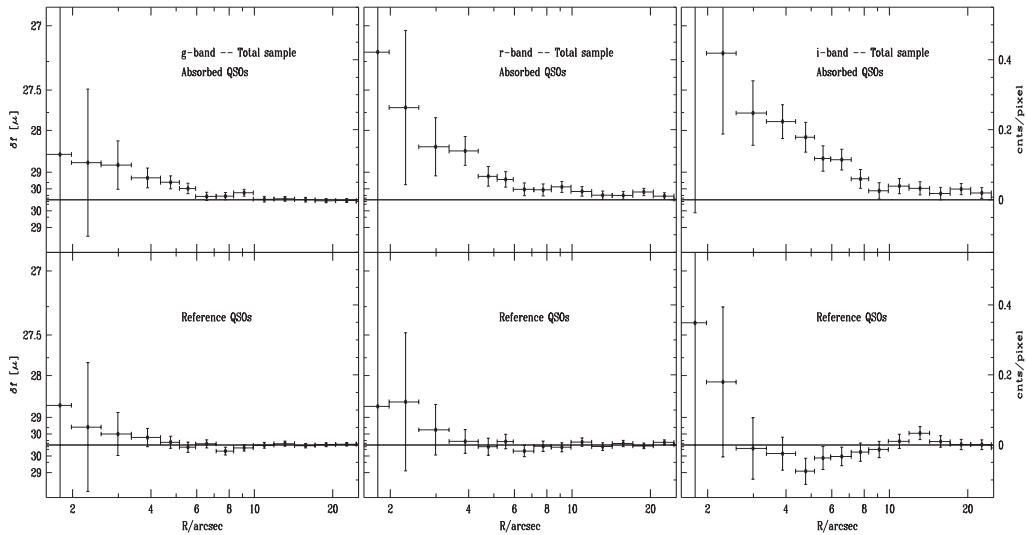}
  \caption{The statistical light present around quasars with (upper
panels) and without MgII absorbers (lower panels) measured in the $g$,
$r$ and $i$-bands after having subtracted the PSF. The light excess
seen around quasars with absorbers corresponds to the light
originating from the absorbing galaxies.}  \label{fig3} \end{center}
\end{figure}

In order to isolate the excess light due to intervening absorbers, we
subtract the mean point spread function (PSF).  The PSF depends on
several parameters: the time of the observation, the position on a
given camera column, the color of the object, etc. In order to build
an effective PSF which is representative of the flux-averaged PSF of
our QSOs extended to large angular distances, we stack stars selected
to match each individual QSO. We first require stars to be observed in
the same observing run and same camera column as the QSO. These stars
are then ranked to minimize a combination of the following quantities:
(i) the difference of Gaussian FWHM with respect to the QSO; (ii) the
angular separation from the QSO; and (iii) the differences in colors with
respect to those of the QSO. To map the PSF with reasonable S/N at
large distances, bright unsaturated stars are also preferred ($r<17.0$
mag is however required). For each QSO in each band, the star that
best fulfills all these criteria is chosen.  To obtain the correct
flux-weighted average PSF, the intensity of each star is rescaled to
the same intensity of the corresponding QSO before stacking.

In Fig.\ref{fig3} we show the surface brightness residuals measured in
the $g$, $r$ and $i$-band after the PSF subtraction for the sample of
810 QSOs with absorbers (upper panels) and for the reference QSO
sample (lower panels). Note that three quasar images could not be used
because of their proximity to a bright star. The stacked PSFs are
normalized with respect to the QSO profiles within an aperture of 4
pixels (1.6\arcsec). The vertical scale is linear in flux intensity
and the horizontal solid line is the background level.  The left
vertical axis displays the surface brightness in mag arcsec$^{-2}$ (in
excess or deficit).  Error bars are computed from the jackknife
covariance matrix and include the budget from the background
uncertainty. Central normalization uncertainties are found to be
negligible. The upper panels show a clear surface brightness
excess. It is detected at 5.5, 7.1 and 9.2$\sigma$ in the $g$, $r$ and
$i$-band, respectively. Such an excess is not seen for the reference
sample. In this case, the results indicate a marginal surface
brightness excess within 4\arcsec, possibly due to the flux
contribution of the quasar host galaxy. Larger datasets will allow us
to measure this effect more precisely.  The excess light that
cross-correlates with the presence of MgII systems provides us with an
estimation of the mean light contribution of the absorbing
galaxies. In addition it should be noticed that the width of this
light excess corresponds to the characteristic size of MgII gas halos
around galaxies.

We have investigated how the signal depends on redshift.  Within
10\arcsec\ the integrated $i$-band absorber magnitude range from
$i=21.50^{+0.09}_{-0.08}$ at $0.37\le z \le0.82$ to roughly two
magnitudes fainter at $0.82\le z \le 2.2$, which is consistent with
the expected dimming as a function of distance. More detailed results
are presented in Zibetti et al. (2005).

Focusing on the redshift range $0.37\le z \le0.82$ we have
investigated the photometric properties of the absorbing
galaxies. Within a 10\arcsec aperture (where the S/N is higher) we
find the average integrated magnitudes to be
$g=23.48^{+0.26}_{-0.20}$, $r=22.27^{+0.13}_{-0.12}$ and
$i=21.50^{+0.09}_{-0.08}$, and the corresponding colors read
$g-r=1.22\pm0.26$ and $r-i=0.77\pm0.15$, indicating a population of
blue galaxies. More detailed results on the implied stellar population
will be presented in a forthcoming paper.

\section{Conclusion}

Using a sample of over 800 MgII systems with $0.4<z<2.2$ we have
presented two statistical analyses constraining the properties of
these absorber systems. After having carefuly selected a population of
reference quasars, we have measured the mean reddening curve induced
by these intervening systems by computing the ratio between the
composute spectra of quasars with and without absorbers.  We do not
find any presence of the 2175 \AA\ bump present in the Milky way and
LMC reddening curves and we find the SMC reddening curve to provide a
good fit to the measured reddening. Assuming the mean MgII absorbers
to have the same metallicity and dust-to-gas ratio as the SMC, we 
measured the mean hydrogen column density of these systems to be
$\sim10^{21}$ atoms cm$^{-2}$.

In addition, we have constrained the mean photometric properties of
MgII absorbing galaxies by measuring the statistical light excess
present around quasars with absorbers. This signal is detected at 5.5,
7.1 and 9.2$\sigma$ in the $g$, $r$ and $i$-band, respectively and
such an excess is not seen for the reference sample. This light
contribution behaves as expected as a function of scale and redshift,
and the measured colors indicate that the mean MgII absorbing galaxies
is a late-type galaxy.

The results presented in this contribution are based on EDR
data. Future analyses will provide much more accurate constraints and
will allow us to investigate the absorber properties as a function of
redshift and for different ion species.

\begin{acknowledgments}

We thank Jim Gunn and Masataka Fukugita for useful
discussions. B.M. acknowledge a support of the Florence Gould
Foundation. Funding for the creation and distribution of the SDSS
Archive has been provided by the Alfred P. Sloan Foundation, the
Participating Institutions, the National Aeronautics and Space
Administration, the National Science Foundation, the U.S. Department
of Energy, the Japanese Monbukagakusho, and the Max Planck Society.

\end{acknowledgments}


\begin{thebibliography}{}

\bibitem[Abazajian et al.(2005)]{dr3} 
{Abazajian, K., et al.} 2005, 
\textit{Astrophysical Journal} 129, 1755 


\bibitem[Bahcall \& Spitzer(1969)]{Bahcall69} 
{Bahcall, J.~N., \& Spitzer, L.~J.} 1969, 
\textit{ApJL} 156, L63 

\bibitem[Bergeron \& Boiss\'e(1991)]{Bergeron91} 
{Bergeron, J., \&  Boisse, P.} 1991, 
\textit{Astronomy \& Astrophysics} 243, 344 

\bibitem[Bertin \& Arnouts(1996)]{Bertin06} 
{Bertin, E., \& Arnouts, S.} 1996, 
\textit{aaps} 117, 393 
 
\bibitem[Boksenberg \& Sargent(1978)]{Bok78} 
{Boksenberg, A., \& Sargent, W.~L.~W.} 1978, 
\textit{ApJ} 220, 42 

\bibitem[Bruzual \& Charlot(2003)]{BC03} 
{Bruzual, G., \& Charlot, S.} 2003, 
\textit{MNRAS} 344, 1000 (BC03)

\bibitem[Churchill et al.(2005)]{Churchill05}
{Churchill, C., Kacprzak, G. G., Steidel, C. C.} 2005, 
\textit{this volume}

\bibitem[Lupton et al. (2001)]{Lupton01} 
{Lupton, R.~H., Gunn, J.~E., Ivezi{\' c}, Z., Knapp, 
G.~R., Kent, S., \& Yasuda, N.} 2001, 
\textit{ASP Conf.~Ser.~238: Astronomical Data Analysis Software and Systems X} 238, 269 

\bibitem[M\'enard et al. (2005)]{menard05}
{M\'enard, B., Nestor, D. B., Turnshek, D. A., Richards, G., Rao, S. M.} 2005,
\textit{in preparation}
 
\bibitem[Nestor et al.(2005)]{nestor05}
{Nestor, D. B., Turnshek, D. A., Rao, S. M.} 2005,
\textit{ApJ in press, astro-ph/0410493}

\bibitem[Schneider et al.(2002)]{Schneider02} 
{Schneider, D.~P., et al.} 2002, 
\textit{Astrophysical Journal} 123, 567 

\bibitem[Stoughton et al.(2002)]{edr} 
{Stoughton, C., et al.} 2002, 
\textit{Astrophysical Journal} 123, 485 

\bibitem[York et al.(2000)]{York00} 
{York, D.~G., et al.} 2000, 
\textit{Astrophysical Journal} 120, 1579 

 \bibitem[Zibetti et al.(2005)]{Zibetti05} Zibetti, S., M{\'e}nard,
B., Nestor, D., \& Turnshek, D.\ 2005, 
\textit{Astrophysical Journal} 120, 1579 

 

\end{thebibliography}
\end{document}